\documentclass[10pt,letterpaper]{article}

\usepackage{opex3}
\usepackage[T1]{fontenc}
\usepackage[latin9]{inputenc}
\usepackage{color}
\usepackage{array}
\usepackage{multirow}
\usepackage{amsmath}
\usepackage{amssymb}
\usepackage{graphicx}
\usepackage{esint}
\usepackage{xargs}[2008/03/08]
\usepackage{floatrow}
\usepackage{cite}

\makeatletter

\pdfpageheight\paperheight
\pdfpagewidth\paperwidth

\providecommand{\tabularnewline}{\\}

\providecommand{\tabularnewline}{\\}

\renewcommand{\maketitle}{}

\makeatother

\begin{document}
\newcommandx\ket[1][usedefault, addprefix=\global, 1=]{|#1\rangle}

\newcommandx\bra[1][usedefault, addprefix=\global, 1=]{\langle#1|}

\newcommandx\avg[1][usedefault, addprefix=\global, 1=]{\langle#1\rangle}

\newcommandx\var[1][usedefault, addprefix=\global, 1=]{\langle(\Delta#1)^{2}\rangle}

\global\long\def\a{\hat{a}}
 \global\long\def\b{\hat{b}}
 \global\long\def\ad{\hat{a}^{\dagger}}
 \global\long\def\bd{\hat{b}^{\dagger}}

\global\long\def\H{\hat{H}}

\global\long\def\I{\mathbb{1}}

\floatsetup[table]{capposition=top}

\title{Hamiltonian design in readout \\
 from room-temperature Raman atomic memory}

\author{Micha\l{} D\k{a}browski, Rados\l aw Chrapkiewicz,$^*$ and Wojciech Wasilewski}
\address{Institute of Experimental Physics, University of Warsaw, Pasteura 5, 02-093 Warsaw, Poland}

\email{$^*$radekch@fuw.edu.pl} 

\maketitle

\begin{abstract}
We present an experimental demonstration of the Hamiltonian manipulation
in light-atom interface in Raman-type warm rubidium-87 vapor atomic
memory. By adjusting the detuning of the driving beam we varied the
relative contributions of the Stokes and anti-Stokes scattering to
the process of four-wave mixing which reads out a spatially multimode
state of atomic memory. We measured the temporal evolution of the
readout fields and the spatial intensity correlations between write-in
and readout as a function of detuning with the use of an intensified
camera. The correlation maps enabled us to resolve between the anti-Stokes
and the Stokes scattering and to quantify their contributions. Our
experimental results agree quantitatively with a simple, plane-wave
theoretical model we provide. They allow for a simple interpretation
of the coaction of the anti-Stokes and the Stokes scattering at the
readout stage. The Stokes contribution yields additional, adjustable
gain at the readout stage, albeit with inevitable extra noise. Here
we provide a simple and useful framework to trace it and the results can be
utilized in the existing atomic memories setups. Furthermore, the shown Hamiltonian
manipulation offers a broad range of atom-light interfaces readily
applicable in current and future quantum protocols with atomic ensembles. 
\end{abstract}


\date{\today}

\section{Introduction}

On-demand retrieval of pre-stored photons can be accomplished using
quantum memories \cite{Lvovsky2009a}. This is an indispensable technique
for long distance quantum communication networks \cite{Duan2001,Kimble2008a}
and for the enhanced generation of multi-photon states \cite{Nunn2013}.
Such states find applications in a linear quantum computing scheme
\cite{Kok2007} or quantum simulators using linear optics \cite{Chiuri2012,Aspuru-Guzik2012}.

Implementations of quantum memories in room-temperature setups are
among the most auspicious in terms of possible future applications due to their
robustness. Until now room-temperature quantum memories have been
realized in solid state systems \cite{Neumann2010,Maurer2012} and
in warm atomic ensembles \cite{Lvovsky2009a} such as gradient echo
memory \cite{Hosseini2011,Higginbottom2012b}, Raman memory \cite{Reim2010,Reim2011}
or EIT memory \cite{Harris1997,Fleischhauer2000a,Lukin2001,Fleischhauer2005a}. 

In this paper we focus on Raman-type atomic memory implemented in
warm rubidium-87 vapors \cite{Bashkansky2012a,Chrapkiewicz2012,Chrapkiewicz2014b}.
In such memory photons are stored in atomic collective excitations
called spin-waves. They are interfaced to photons via off-resonant
Raman transitions. Storage times up to 30 ms in a single spin-wave
mode was demonstrated \cite{Jensen2010}. Multimode storage, both
temporal \cite{Hosseini2011} and spatial \cite{Higginbottom2012b,Chrapkiewicz2012},
is also feasible. With the use of multiple transverse spin-wave modes
images can be stored and retrieved \cite{Vudyasetu2008,Firstenberg2013}.
The storage time of the multimode memory is limited by diffusional
decoherence and thus can be prolonged by increasing the beam size
\cite{Chrapkiewicz2014b,Firstenberg2013} at the expense of laser
intensity.

Here we demonstrate experimentally the theoretical concept of ``Hamiltonian
design'' proposed in \cite{DeEchaniz2008}. We implement the Hamiltonian
design at the readout from Raman atomic memory. Ideal readout from
Raman atomic memory relies on a pure anti-Stokes scattering process
which maps the spin-wave state onto photons state. In real systems
additional Stokes scattering at the readout is virtually inevitable
and it is a source of extra noise \cite{Reim2011,Michelberger2014}.
By modifying the interaction Hamiltonian we are able to control the
relative contributions of anti-Stokes and Stokes scattering processes.
In the particular setting of our experiment this enables parametric
amplification of the readout, albeit with extra noise. 

In this paper we use simple theoretical model describing the coaction
between anti-Stokes and Stokes scattering in readout. Previous approaches
to describe theoretically the readout of spin-wave excitations in
Raman scattering were focused only on pure anti-Stokes process.
Theoretical descriptions of readout from Raman-type memory have already
been given for a number of cases, including spatio-temporal evolution
with losses \cite{Raymer2004}, temporal eigen-modes \cite{Wasilewski2006},
optimized retrieval \cite{Gorshkov2007d}, and spatial modes \cite{Zeuthen2011}.
However, in those papers emphasis was put on the pure anti-Stokes
scattering. In the present paper we use the extended description of
the readout process \cite{Wasilewski2009}, taking into account the
four-wave mixing that includes the Stokes scattering. 

In the experiment we detect scattered light with spatial resolution
and temporal gating. This enables directly relating the experimental
results to the theoretical predictions for temporal evolution of scattered
light. Stokes and anti-Stokes light contributions can be distinguished
through intensity correlation measurements and quantified via careful
post-processing.

We provide evidence for
a possibility of engineering a wide range of atom-light interfaces
which can be described theoretically as simultaneous readout and parametric
amplification. This may enable enriching purely optical quantum metrology
\cite{Vitelli2010} or communication \cite{Fossier2009} schemes with
quantum storage capabilities.

This paper is organized as follows: In Sec. 2 we introduce a single-mode,
theoretical model of temporal evolution of Stokes and anti-Stokes
in readout, Sec. 3 describe experimental details, in Sec. 4 we present
the phenomenological model utilized for data analysis and the results;
finally Sec. 5 concludes the paper.

\section{Theory}

\begin{figure}[b]
\includegraphics[width=0.67\textwidth]{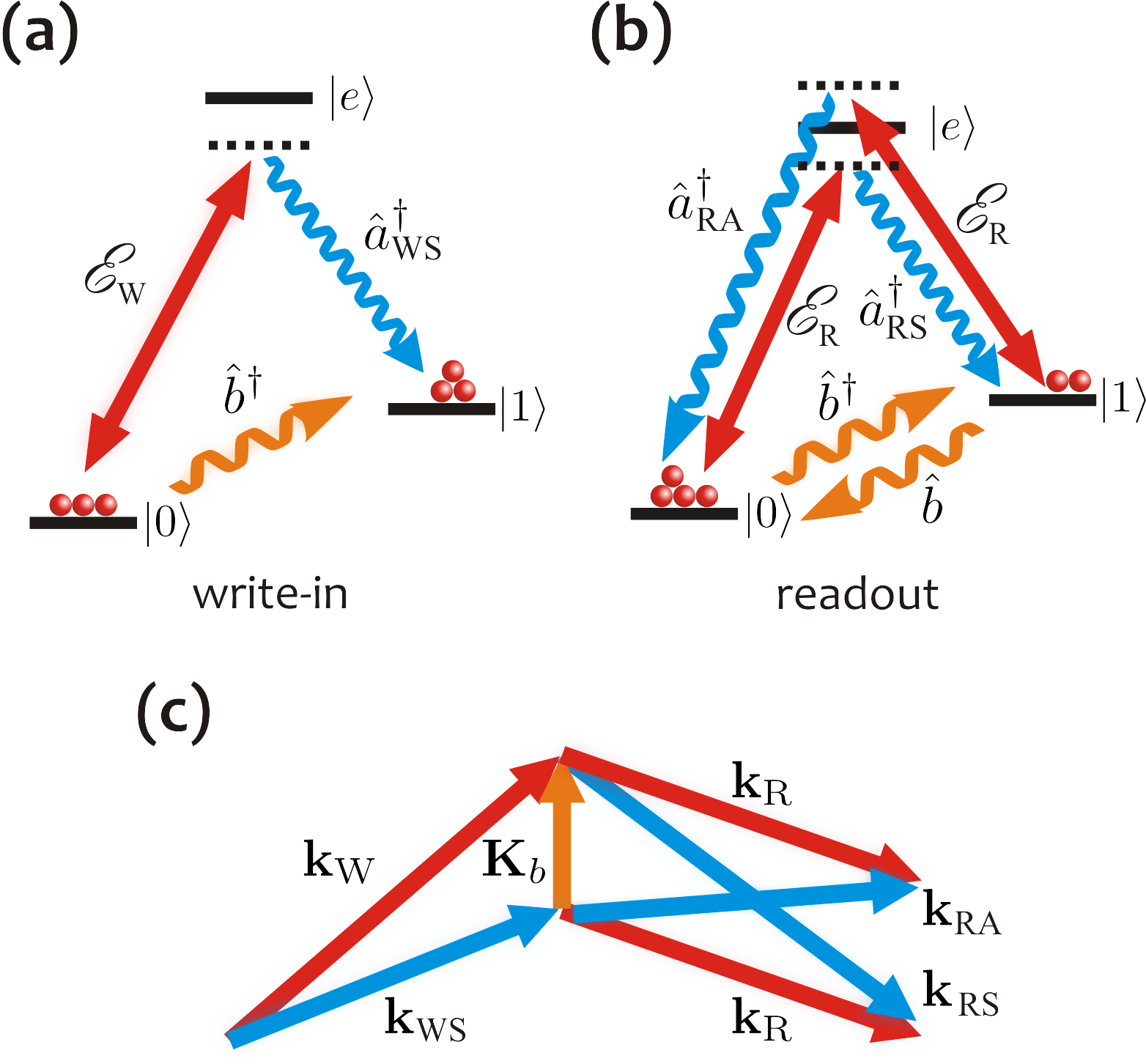}\centering

\protect\protect\caption{Atomic levels and phase matching in $\Lambda$-scheme Raman scattering
induced by classical laser field $\mathcal{E}_{\mathrm{W}}$ and $\mathcal{E}_{\mathrm{R}}$.
(a) In the spontaneous write-in process Stokes photons ($\protect\ad_{\mathrm{WS}}$
- mode) and spin-wave excitation (\textbf{$\protect\bd$ -} mode)
are created pairwise. (b)\textbf{ }Four-wave mixing in readout consists
in simultaneous anti-Stokes and Stokes scattering into modes $\protect\ad_{\mathrm{RA}}$
and $\protect\ad_{\mathrm{RS}}$. (c) Phase matching condition or
momentum conservation dictates the wave vectors of single photons
coupled to a spin-wave excitation with a certain wave vector $\mathbf{K}_{b}$.
Write beam with $\mathbf{k}_{\mathrm{W}}$ wave vector is scattered
as Stokes photon with $\mathbf{k}_{\mathrm{WS}}$ wave vector, while
read beam with $\mathbf{k}_{\mathrm{R}}$ wave vector either scatters
Stokes or couples anti-Stokes of the respective wave vectors $\mathbf{k}_{\mathrm{RS}}$,
$\mathbf{k}_{\mathrm{RA}}$.\label{fig:poziomy-dopasowaniefazowe}}
\end{figure}

We begin by synthesizing a simple, plane-wave theoretical model of
the readout by four-wave mixing from a Raman memory to qualitatively
support our experimental results. The spin-wave excitation in the
atomic medium can be created by different means. Light can be absorbed
in a two-photon Raman transition to store previously prepared state
of light \cite{Reim2011,Michelberger2014}. In our experiment we induced
spontaneous Stokes scattering using write pulse with a wave vector $\mathbf{k_{\mathrm{W}}}$
to populate the spin-waves, i.e. collective atomic excitations from
the levels $\ket[0]$ to $\ket[1]$ as in Fig. \ref{fig:poziomy-dopasowaniefazowe}(a).
Both approaches results in the spin-wave excitation, however the latter
which we use does not require special efforts to match the photons
from the external source to the memory bandwidth. Ideally, the number
of created spin-wave excitations $n_{b}$ with a certain wave vector
$\mathbf{K}_{b}$ equals the number of scattered Stokes photons with
wave vector $\mathbf{\mathbf{\mathbf{k}_{\mathrm{WS}}}=k_{\mathrm{W}}-K}_{b}$.
We were able to estimate those numbers in each single iteration of
the experiment.

Here we focused on the retrieval stage at which the spin-wave excitations
are converted to photons in four-wave mixing induced by the read laser
pulse depicted in Fig.\ref{fig:poziomy-dopasowaniefazowe}(b). The
read laser is assumed to be plane-wave with a wave vector $\mathbf{k}_{R}$.
Spin-wave with a certain wave vector $\mathbf{K}_{b}$ are coupled
to anti-Stokes and Stokes fields with wave vectors $\mathbf{k}_{\mathrm{RA}}=\mathbf{k}_{\mathrm{R}}+\mathbf{K}_{b}$
and $\mathbf{k}_{\mathrm{RS}}=\mathbf{k}_{\mathrm{R}}-\mathbf{K}_{b}$
respectively. In the experiment those weak light fields illuminated
distinct pixels of the camera which was located in the far field.
They were also shifted with respect to the initial Stokes photons
with the wave vector $\mathbf{\mathbf{k}_{\mathrm{WS}}=k_{\mathrm{W}}-K}_{b}$
due to different direction of the write beam $\mathbf{k}_{\mathrm{W}}$.
We summarize the phase matching condition relating the wave vectors
in Fig. \ref{fig:poziomy-dopasowaniefazowe}(c).

To describe the interaction at the readout stage we use bosonic operators
of the weak field light modes $\a_{\mathrm{RA}}$ and $\a_{\mathrm{RS}}$.
The spin-wave can be described with the Holstein-Primakhoff approximation
by a bosonic annihilation operator $\hat{b}$ which removes one excitation
from spin-wave with the wave vector $\mathbf{K}_{b}$ \cite{Koodynski2012b}.
The Hamiltonian describing both the Stokes and anti-Stokes scattering
at the readout stage can be obtained by eliminating the excited level
adiabatically \cite{Krauter2013,Wasilewski2009}: 
\begin{equation}
\H_{\mathrm{R}}=i\hbar\chi\ad_{\mathrm{RA}}\b+i\hbar\xi\ad_{\mathrm{RS}}\bd+H.c.\label{eq:HR}
\end{equation}
where $\chi$ and $\xi$ are the coupling coefficients for anti-Stokes
and Stokes Raman transitions \cite{Raymer1981}. The coupling coefficients
squared $\chi^{2}$ or $\xi^{2}$ can be calculated \cite{Gorshkov2007}
as $\Gamma d\Omega_{p}^{2}/\Delta^{2}$ where $d$ is optical depth
at resonance, $\Omega_{p}$ is the pump Raman frequency and $\Delta$
is the detuning of the pump from transition from level 1 or 0 to excited
state for anti-Stokes coupling $\chi^{2}$ or $\xi^{2}$respectively.
Such a Hamiltonian (Eq. \eqref{eq:HR}) has been used in several papers
describing experiments on room-temperature atomic vapors, i.e. teleportation
between distant atomic objects \cite{Krauter2013}.

The above Hamiltonian leads to coupled Maxwell-Bloch equations, which
can be integrated as e.g. in \cite{Wasilewski2009,Koodynski2012b}.
This yields input-output relations linking the operators of the incoming
light fields and the initial spin-wave state to the outgoing fields
and the final spin-wave state.

We used the input-output relations to compute the quantities observed
in the experiments, that is the mean number of scattered photons as
a function of time in anti-Stokes $\avg[\ad_{\mathrm{RA}}(t)\a_{\mathrm{RA}}(t)]$
and Stokes fields $\avg[\ad_{\mathrm{RS}}(t)\a_{\mathrm{RS}}(t)]$
respectively. With the assumption that the incoming light fields are
in a vacuum state and the mean number of initial spin-wave excitations
equals $n_{b}=\avg[\bd(0)\b(0)]$, the result is: 
\begin{equation}
\avg[\ad_{\mathrm{RA}}(t)\a_{\mathrm{RA}}(t)]=\underbrace{{\chi^{2}e^{t\left(\xi^{2}-\chi^{2}\right)}}}_{G_{\mathrm{RA}}(t)}n_{b}+\underbrace{{\frac{\chi^{2}\xi^{2}}{\xi^{2}-\chi^{2}}(e^{t\left(\xi^{2}-\chi^{2}\right)}-1)}}_{S_{\mathrm{RA}}(t)},\label{eq:I1}
\end{equation}
\begin{equation}
\avg[\ad_{\mathrm{RS}}(t)\a_{\mathrm{RS}}(t)]=\underbrace{\xi^{2}e^{t\left(\xi^{2}-\chi^{2}\right)}}_{G_{\mathrm{RS}}(t)}n_{b}+\underbrace{\frac{\xi^{2}}{\xi^{2}-\chi^{2}}(\xi^{2}e^{t\left(\xi^{2}-\chi^{2}\right)}-\chi^{2})}_{S_{\mathrm{RS}}(t)},\label{eq:I2}
\end{equation}

\begin{figure}[b]
\begin{centering}
\includegraphics[width=0.90\textwidth]{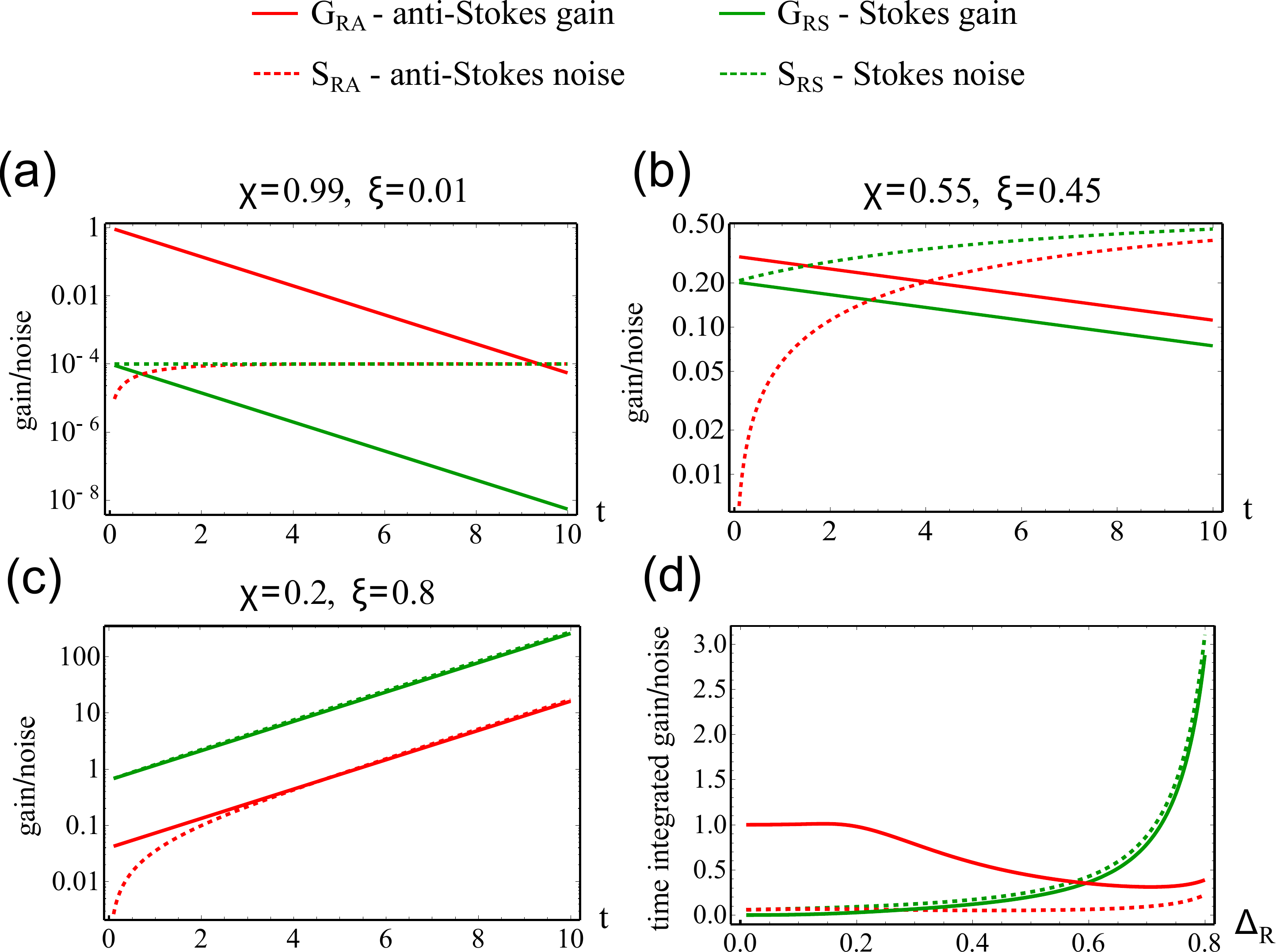} 
\par\end{centering}

\protect\protect\caption{(a) to (c) - temporal evolution of gains $G_{i}$ and spontaneous
noises $S_{i}$ building up the light fields. Data plotted for different
values of coupling coefficients $\chi$ and $\xi$ correspond to detuning
parameters $\Delta_{R}$= 0.3, 0.6 and 0.8. (d) Integrated readout
gains $\bar{G}_{i}$ and spontaneous noises $\bar{S}_{i}$ versus
detuning $\Delta_{R}$.\label{fig:TeoriaOdOdstrojenia}}
\end{figure}

The underbraced terms $G_{\mathrm{RA}}(t)$ and $G_{\mathrm{RS}}(t)$
represent the readout gains, that is the number of scattered photons
per each input spin-wave excitation. In turn, the terms $S_{\mathrm{RA}}(t)$
and $S_{\mathrm{RS}}(t)$ represent the spontaneous scattering into anti-Stokes and Stokes quantum fields, which is independent of the initial state of quantum fields. Just
as expected, they appear only in the presence of Stokes scattering
$\xi\neq0$. Note that both coupling coefficients $\chi$ and $\xi$
appear in both formulas (\ref{eq:I1}) and (\ref{eq:I2}), confirming
that anti-Stokes and Stokes scattering cannot be treated separately
and should be viewed together as a full four-wave mixing process.

Note that the above results were derived neglecting all decoherence. In the experiment, the most important sources of decoherence are diffusion of atoms and spontaneous emission from the excited state in random direction. Both can be neglected provided the optical depth is large and the duration of the interaction short.

Let us proceed to examining the evolution of scattered anti-Stokes
and Stokes light in a few typical cases.

For $\xi\simeq0$ there is virtually no Stokes scattering while the
anti-Stokes scattering appears instantaneously and decays exponentially
as depicted in Fig. \ref{fig:TeoriaOdOdstrojenia}(a). There is virtually
no spontaneous noise $S_{\mathrm{RA}}\simeq0$ and the time integrated
gain reaches unity $\bar{G}_{\mathrm{RA}}=\int G_{\mathrm{RA}}\mathrm{d}t=1$.
In the ideally case of $\xi=0$ this leads to the readout Hamiltonian
of form $\H_{\mathrm{R}}\sim\ad_{\mathrm{RA}}\b+H.c.$ which represent
purely anti-Stokes scattering process.

In real systems Stokes interaction is typically unavoidable. The case
where $\chi>\xi>0$ is depicted in Fig. \ref{fig:TeoriaOdOdstrojenia}(b).
Here the readout appears in both the anti-Stokes and Stokes fields
with comparable intensity and decays over time. Time-integrated gain
may reach over unity $\bar{G}_{i}=\int G_{i}\mathrm{d}t$ > 1, yet with
spontaneous noise that slowly builds up $S_{i}>0$, for $i=\mathrm{RA,\, RS}$.
This is the setting utilized in \cite{Boyer2008e}.

A situation where Stokes interaction dominates, $\xi>\chi$, is depicted
in Fig. \ref{fig:TeoriaOdOdstrojenia}(c). In this case readout goes
predominantly to Stokes field with exponentially increasing gain.
Yet, the noise intensity virtually equals the gain $G_{i}\simeq S_{i}$
for both fields. In the limit of $\xi\gg\chi$ we can put $\chi=0$
which leads to the readout Hamiltonian $\H_{\mathrm{R}}\sim\ad_{\mathrm{RS}}\bd+H.c.$
consists only the contribution of Stokes scattering process.

It is instructive to integrate gains $G_{\mathrm{RA}}(t)$ and $G_{\mathrm{RS}}(t)$,
and noises $S_{\mathrm{RA}}(t)$ and $S_{\mathrm{RS}}(t)$ over the
time of interaction and inspect them as functions of coupling coefficients.
Coupling coefficients $\chi$ and $\xi$ are inversely proportional
to the detuning of the read laser from the excited level $\Delta_{R}$
\cite{Raymer1981,Koodynski2012b}. One coupling coefficient can be
increased at the expense of the other by tuning the frequency of the
read laser when it is in between the $\ket[0]\leftrightarrow\ket[e]$
and $\ket[1]\leftrightarrow\ket[e]$ resonances as in Fig. \ref{fig:poziomy-dopasowaniefazowe}(b).

In Fig. \ref{fig:TeoriaOdOdstrojenia} (d) we plot the time integrated
gains $\bar{G}_{i}=\int G_{i}\mathrm{d}t$ and noises $\bar{S}_{i}=\int S_{i}\mathrm{d}t$
as a function of the detuning $\Delta_{R}$ from $\ket[0]\leftrightarrow\ket[e]$
transition in the units of ground state splitting. We assumed $\chi\propto1/\Delta_{R}$
and $\xi\propto1/(1-\Delta_{R})$. Different characters of anti-Stokes
and Stokes interaction causes asymmetry of the plot in Fig. \ref{fig:TeoriaOdOdstrojenia}
(d). The integrated readout gain in the anti-Stokes domination regime
$\chi\gg\xi$ remains equal to unity $\bar{G}_{RA}=1$ and the noise
is suppressed $\bar{S}_{RA}\ll1$ . These are the conditions for perfect
unamplified readout.

On the contrary, in the Stokes domination regime $\chi\ll\xi$ the
integrated gain $\bar{G}_{RS}$ varies almost exponentially with the
Stokes coupling $\xi$ entailing elevated noise $\bar{S}_{RS}$. In
this domain four-wave mixing enhances the anti-Stokes emission and
$\bar{G}_{RA}$ rises although the relative contribution $\bar{G}_{RA}/\bar{G}_{RS}$
diminishes. Note that this is the regime we use to populate the spin-waves
by inducing spontaneous Stokes scattering with write pulse.

Finally, let us consider a single realization of the four-wave mixing
process. Assume that the number of initial spin-wave excitations $n_{b}$
has been measured by counting the Stokes photons $n_{WS}$ scattered
during prior write-in, ideally $n_{WS}=n_{b}$. In this particular
iteration the number of photons in the anti-Stokes and the Stokes
field equals 
\begin{equation}
n_{i}=\bar{G}_{i}n_{b}+\check{S}{}_{i},\quad i=\mathrm{RS,\, RA}\label{eq:photon-gain-relation}
\end{equation}
where $\check{S}_{i}$ is the noise treated here as an independent
random variable. $\check{S}_{i}$ is the time integrated spontaneous
scattering and its mean is calculated based on Eqs. (\ref{eq:I1}),
(\ref{eq:I2}) $\avg[\check{S}_{i}]=\bar{S}_{i}$. Later on we refer
to the correlation of the number of anti-Stokes $n_{\mathrm{RA}}$
or Stokes photons $n_{\mathrm{RS}}$ with the initial spin-wave excitation
$n_{b}$. The above formula links the correlated part to the integrated
readout gain $\bar{G}_{i}$ and the uncorrelated noise to the integrated
spontaneous contribution $\bar{S}_{i}$.

\section{Experiment}

\begin{figure}[b]
\begin{centering}
\includegraphics[width=0.8\textwidth]{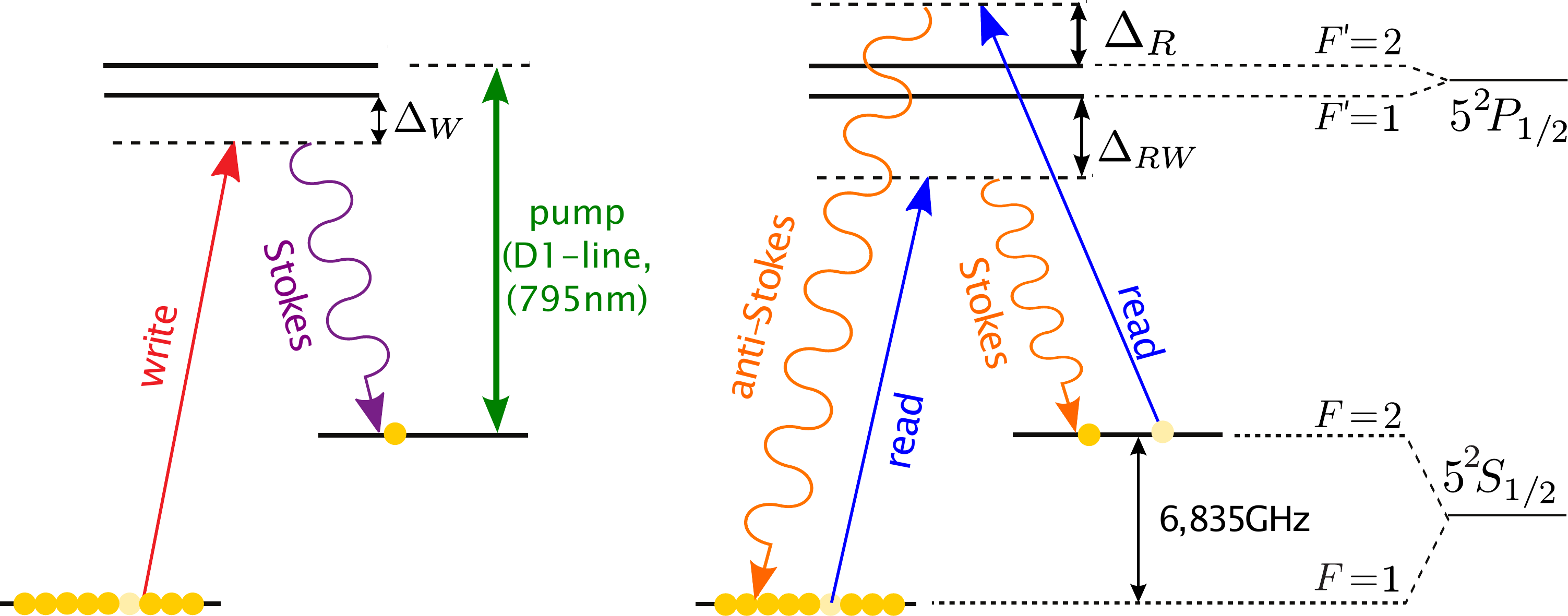} 
\par\end{centering}

\protect\protect\caption{Configuration schemes of \protect\protect\textsuperscript{87}Rb
hyperfine energy levels at D1 line (795nm) using write-in and readout
respectively. Laser beams and scattered light detunings in both processes are also depicted.\label{fig:poziomy-eksperyment}}
\end{figure}

We implemented Raman-type atomic memory in rubidium-87 vapors. We
applied spontaneous write-in process into memory similarly as in previous
works \cite{VanderWal2003,Bashkansky2012a}. The spontaneous write-in
process had been proven to efficiently create the spin-wave excitations,
thus we could focus on the retrieval characteristics. Here we extended
the setup and measurements schemes applied and explained in detail
in our previous papers \cite{Chrapkiewicz2012,Chrapkiewicz2014b}.

We used a 10 cm glass cell containing pure $^{87}$Rb isotope with
krypton as a buffer gas under pressure of 1 torr. The cell was heated
by bifilar windings to 90$^{o}$C equivalent to the optical density
of $135$. The cell was magnetically shielded to avoid decoherence
produced by stray magnetic field and the main source of decoherence
was due to diffusion \cite{Chrapkiewicz2014b}.

We used three external cavity diode lasers: pump, write and read laser,
operating at D1 line of $^{87}$Rb (795 nm). \textcolor{black}{The
frequencies of all laser beams and the scattered light are sketched
against the $^{87}$Rb level scheme in Fig. \ref{fig:poziomy-eksperyment}.
}The pump laser was resonant to F=2 $\rightarrow$F'=2 transition,
while the write and read laser were detuned to the red from transition
F=1$\rightarrow$F=1'. The detuning of the write laser was \textcolor{black}{$\Delta_{\mathrm{W}}=1.77\ \mathrm{GHz}$
and it was} kept constant throughout all measurements while the detuning
of the read laser $\Delta_{R}$ varied. The pump and write lasers were
frequency locked using the DAVLL setup \cite{McCarron2007}. The exact
values of the write and read lasers detunings were set repeatedly
and measured precisely using Doppler-free saturated absorption spectroscopy
inside an auxiliary rubidium cell and their reference beat-note signal
was measured on a fast photodiode. Inside the cell write, read and
pump lasers had the power of 6.8 mW, 4.5 mW and 75 mW respectively.
During the experimental sequence the lasers power fluctuated at a level of 5\% and the frequencies were changing by no more than 50MHz.

\begin{figure}[b]
\begin{centering}
\includegraphics[width=0.8\textwidth]{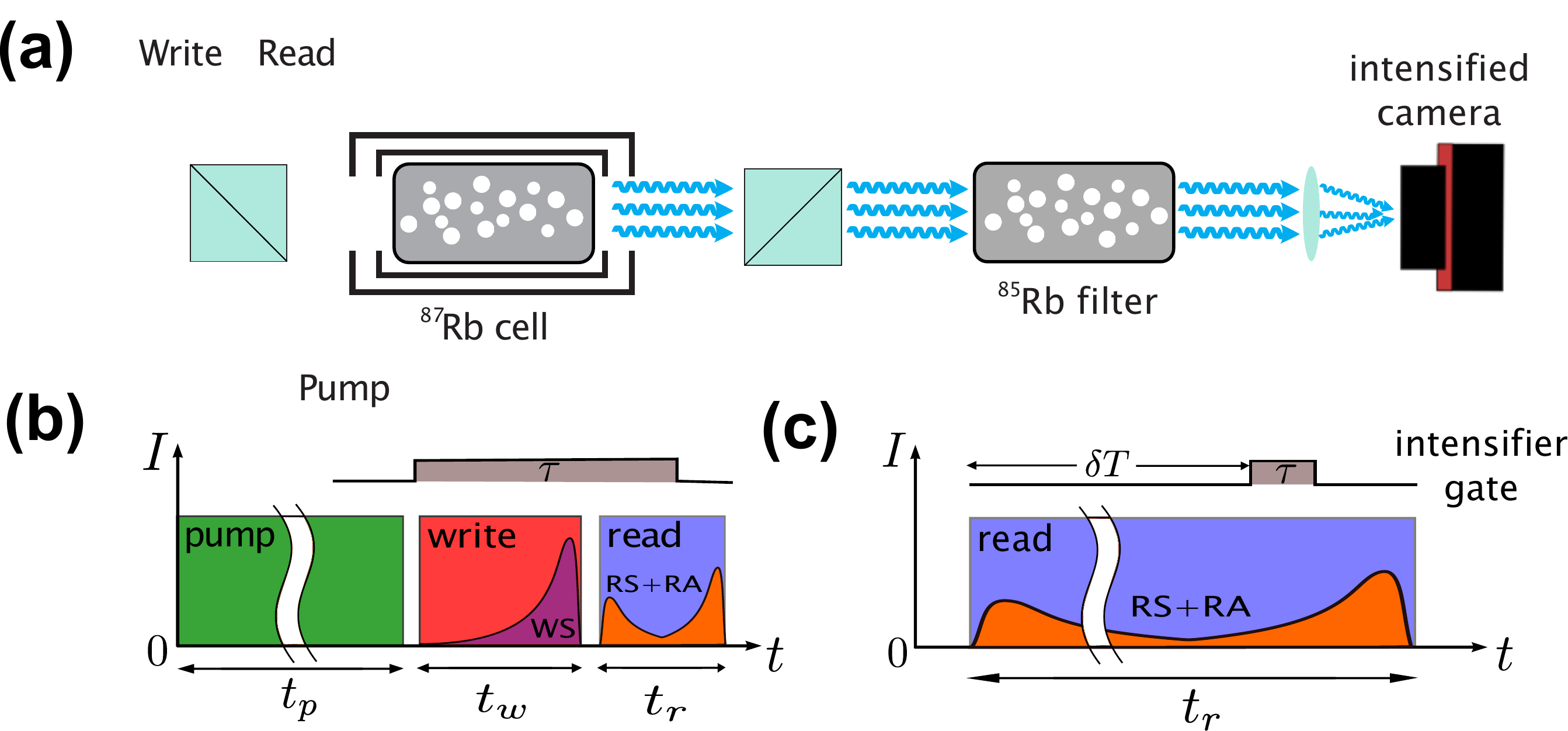} 
\par\end{centering}

\protect\protect\caption{(a) Experimental setup: write and read laser beams propagate forward
to the intensified camera and the pump beam propagates backward. \protect\protect\textsuperscript{87}Rb
- atomic memory cell, \protect\protect\textsuperscript{85}Rb - absorption
filter, blue arrows correspond to scattered light. (b) Pulse sequence
with long intensifier gate covers the whole write and a part of the
read pulse (both rectangular). Exponential shapes in front of write
and read pulses are typical time-resolved intensities of the scattered
light observed on the intensified camera, WS and RS stand for Stokes
while RA for anti-Stokes scattering respectively. (c) Scheme of the
measurement of the temporal evolution of the readout light using short
gate duration $\tau=250$ ns.\label{fig:setup}}
\end{figure}

The simplified schematic of the setup is shown in Fig. \ref{fig:setup}(a).
All of the three horizontally polarized beams overlapped inside the cell
where their diameters were 6 mm for the pump beam and 4 mm for both
write and read beams. The write and read beams were tilted at the angle
$\theta=2\ \mathrm{mrad}$. Acousto-optic modulators were used to
shape rectangular pulses with rising time of $1\ \mu\mathrm{s}$ from
laser beams. The pulse sequence applied in the experiment is depicted
in Fig. \ref{fig:setup}(b). The sequence was initiated by optical
pumping of rubidium atoms into the ground state of \textsuperscript{87}Rb
S$^{1/2}$ F=1. The 700~$\mu$s long rectangular pulse of 75 mW power
yielded a pumping efficiency of 98\%. Then we applied a $10\ \mu\mathrm{s}$
long rectangular write pulse to create spin-wave excitations between
F=1 and F=2 levels together with the Stokes scattering. The rectangular
read pulse started right after the end of the write pulse and its
duration was set to $40\ \mu\mathrm{s}$. The read pulse generated
both anti-Stokes and Stokes scattering. The total time duration of
write and read pulse corresponds to the average atomic displacement
of c.a. 0.5 mm due to diffusion \cite{Chrapkiewicz2014b} which was
small enough to keep the atoms in the range of the pump beam size. 

We separated the horizontally polarized laser beams from the vertically
polarized scattered fields by polarization and spectral filtering,
which yielded the total attenuation factor for laser beams that exceeded
$10^{9}$. For spectral filtering we applied a $^{85}$Rb absorption
filter \textcolor{black}{at $130^{\circ}$C }placed in magnetic field
which increase the absorption by broadening and shifting the $^{85}$Rb
spectral lines.\textcolor{black}{{} The attenuation of laser beams inside
the 30 cm length absorption filter was at least $10^{4}$. We also
measured the transmission of the filter at frequencies corresponding
to the scattered light generated at write-in and readout stages. They
were found to be 12\% for Stokes scattering in write-in and 76\% for
all applied frequencies of anti-Stokes and Stokes in readout. }

\begin{figure}[t]
\begin{centering}
\includegraphics[width=0.68\textwidth]{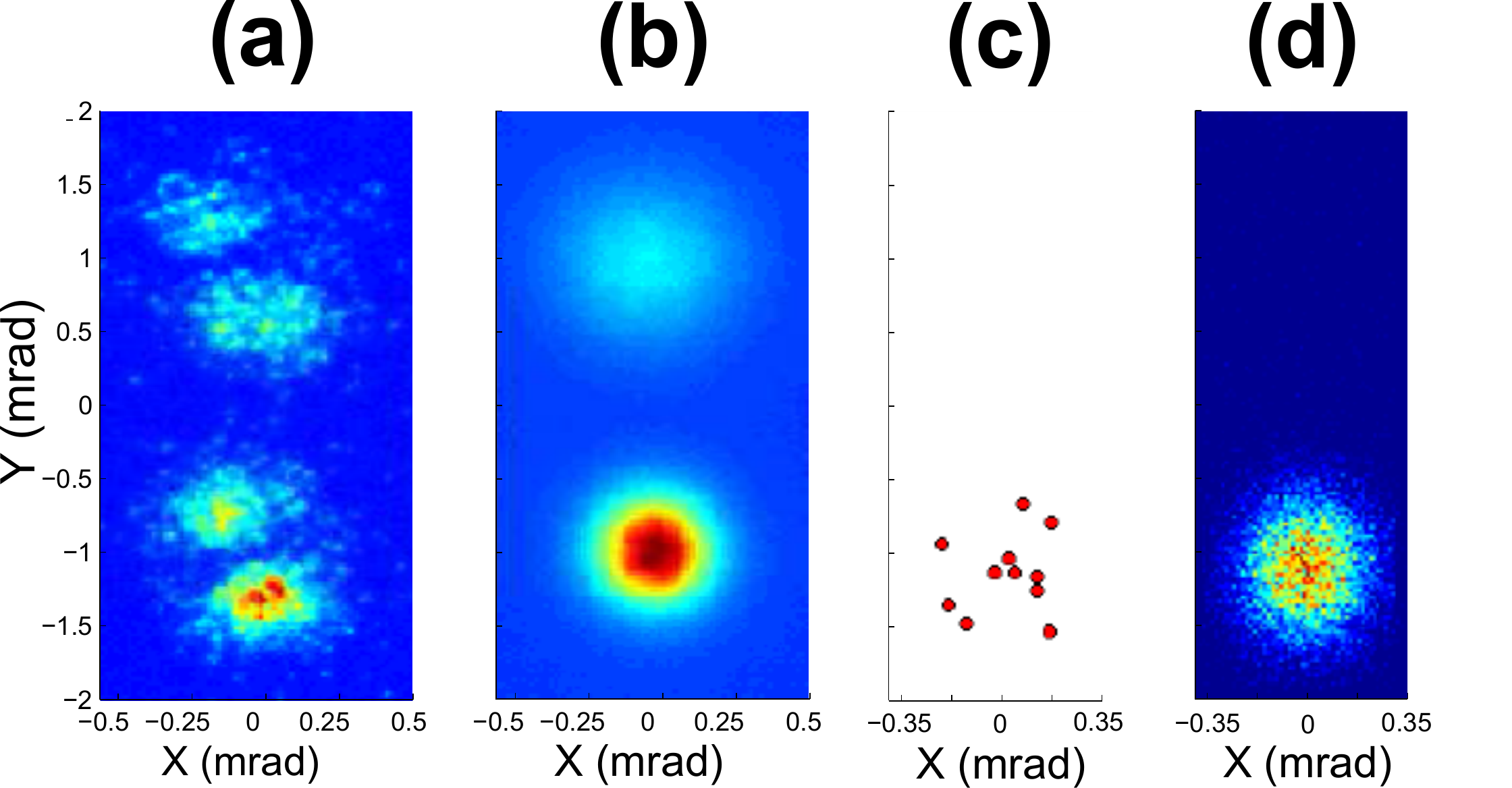} 
\par\end{centering}

\protect\protect\caption{Representative images of the retrieved field, write-in and read-out in upper and lower parts respectively. (a) Intensity map in a single shot obtained using a long gate in the linear regime of camera operation and (b) the average over $10^4$ frames. (c) Photon positions in a single shot \cite{Chrapkiewicz2014aa} obtained using a short gate positioned in the readout stage and (d) the total number of photons per sCMOS pixel summed over 2000 frames. \label{fig:images}}
\end{figure}

In our system we generated and retrieved spatially multimode light
as described in detail in \cite{Chrapkiewicz2012,Koodynski2012b}.
This light was detected in the far field by a gated image intensifier
coupled to sCMOS camera \cite{Chrapkiewicz2014aa}. We used the intensified
camera in two different operational schemes. In the first scheme we
applied a long gate pulse which covered both the write and the read
pulse as in Fig. \ref{fig:setup}(b). We set the gain of the  intensifier
to a low value. Then the response of the camera system was proportional
to the intensity of scattered light. We calibrated the excess noise
contributed by the image intensifier and made sure that it was insignificant
as compared to the shot-to-shot intensity fluctuations of scattered
light. In the second scheme we used a short, delayed gate of $\tau=250$
ns duration as depicted in Fig. \ref{fig:setup}(c). Here we set the
image intensifier to high gain and the camera system was sensitive
to single photons \cite{Chrapkiewicz2014aa}. We also utilized an
intensified camera as a photon number resolving detector \cite{Chrapkiewicz2014bb}
and we counted all photons with a quantum efficiency of 20\% in the
region of readout scattering. Representative images: single shots and averaged intensities obtained typically in
those regimes are depicted in Fig. \ref{fig:images}. Note that in Fig. 5(a) the number and localization of speckles are random so they average to the smooth intensity profile as shown in Fig. 5(b). Both measurement regimes
are phase insensitive although spatially resolved, homodyne-type detection
was also reported \cite{Boyer2008e} and then it could be used to
directly measure phase noise and squeezing properties of the generated
light. 

\section{Results}

\subsection{Temporal evolution of readout intensity}

\begin{figure}[t]
\begin{centering}
\includegraphics[width=0.55\textwidth]{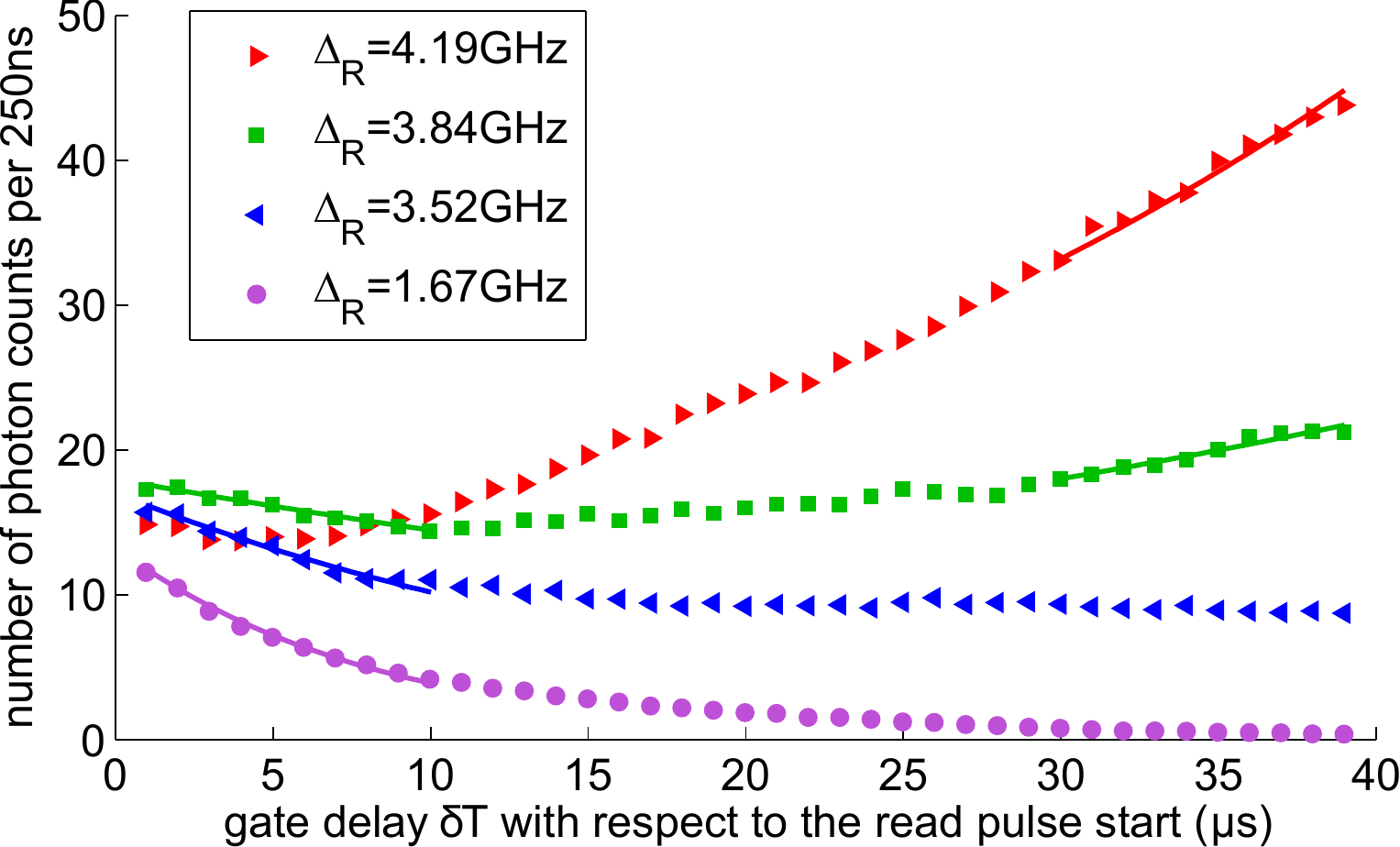} 
\par\end{centering}

\protect\protect\caption{Average number of photon counts detected at a large solid angle around
read beam per a gate duration of 250 ns for different readout laser
detunings $\Delta_{R}$ from \protect\protect\textsuperscript{87}Rb
F=2 $\rightarrow$ F'=2 resonance. The curves represent exponential
fits to the first and last data points. The size of errorbars is comparable
with marker size.\label{fig:TimeEvolution}}
\end{figure}

At first we measured the temporal evolution of the mean number of
photons emitted during readout stage. For this purpose we counted
the scattered photons using an intensified camera with a gate duration
of $\tau=250$ ns. We collected all photons in the specified circular
camera region around the point corresponding to the center of the
read beam i.e. wave vector $k_{\mathrm{R}}$. Thus, we measured the
anti-Stokes and the Stokes scattering together, both their spontaneous
and stimulated parts. In Fig. \ref{fig:TimeEvolution} we present
the results of measurement for mean number of photons versus gate
delay with exponential fits. Each data point was obtained by averaging
2000 frames in the area of angular diameter $0.5\ \mathrm{mrad}$
and the time separation between the data points was 1 $\mu$s. The
sequence was repeated for different detunings $\Delta_{R}$ of the
read beam from F=2$\rightarrow$F'=2 resonance.

The results plotted in Fig. \ref{fig:TimeEvolution} vividly depict
a transition from decay to growth of the readout scattered light intensity.
For a small detuning $\Delta_{R}=1.67$~GHz the read beam is close
to the F=2$\rightarrow$F'=2 resonance and we expected the anti-Stokes
coupling to dominate. Indeed, we observed an exponential decay as
predictied. At the opposite extreme $\Delta_{R}=4.19$~GHz, in turn,
the read beam is much closer to the F=1$\rightarrow$F'=1 transition
and here it was the Stokes field that we expected to dominate. Again,
the total number of scattered photons increases as expected, compare
Fig. \ref{fig:TeoriaOdOdstrojenia}(c). Nonetheless, in the experiment
the growth starts after the first $4$ $\mu$s of the evolution during
which the number of scattered photons slightly decreases. In the intermediate
cases we observed initial decay followed by growth. These cases might
be associated with the case presented in Fig. \ref{fig:TeoriaOdOdstrojenia}(b),
where stimulated field contributions decay while the spontaneous contributions
rise. Moreover, in the observed evolution the spin-wave decoherence
certainly plays an important role, which, however, is not accounted
for by our simplified model.

\begin{figure}[t]
\begin{centering}
\includegraphics[width=0.65\textwidth]{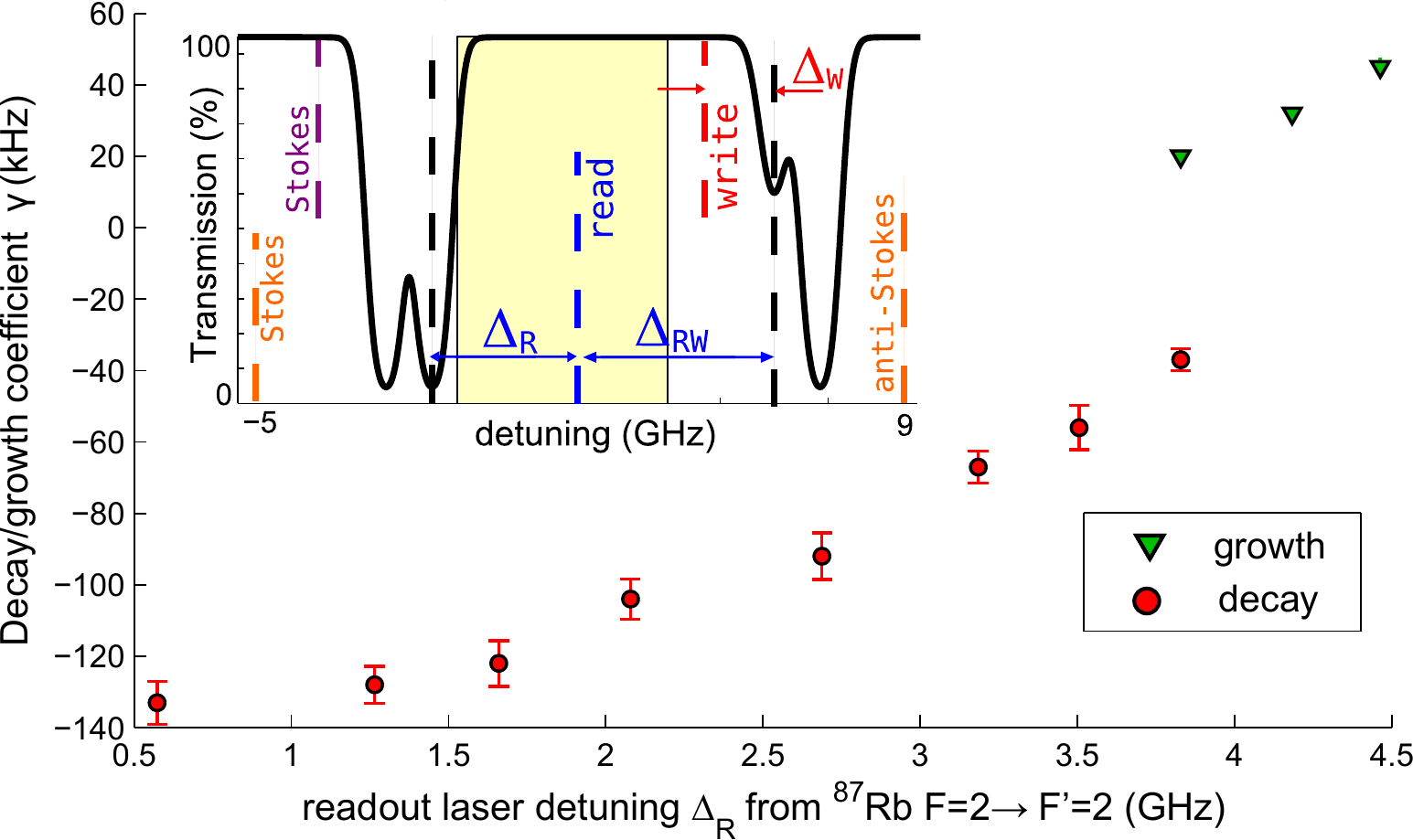} 
\par\end{centering}

\protect\protect\caption{Exponential decay and growth coefficients for scattered light found
by fitting data as in Fig. \ref{fig:TimeEvolution} versus read laser
detuning $\Delta_{R}$. For triangle points the size of errorbars
are smaller than the marker size. Inset: \protect\protect\textsuperscript{87}Rb
D1 line absorption spectrum with read and write laser frequencies
marked.\label{fig:Decay-Growth-Coeffs}}
\end{figure}

In Fig. \ref{fig:Decay-Growth-Coeffs} we plot decay and growth coefficients
as functions of detuning $\Delta_{R}$ of the read beam from F=2$\rightarrow$F'=2
resonance. The coefficients were calculated by fitting exponential
decay or growth to the first or the last {10~$\mu$s},
as depicted in Fig. \ref{fig:TimeEvolution}.
Fitting entire 40~$\mu$s with a single exponential fails because
the significant influence of decoherence and multimode character of
scattering on this time scale is not capured by our simple thoeretical
model. The dependence of the growth and decay coefficients on the detuning $\Delta_{R}$
agree qualitatively with predictions of the theoretical model as they
rise along with $\Delta_{R}$. However, there is a discontinuity between
the growth and decay coefficients at $\Delta_{R}=3.8$~GHz that may
be attributed to spin-wave decoherence influence.

\subsection{Spatial resolving of anti-Stokes and Stokes scattering}

The analysis of intensity correlations between light scattered in
various directions provides valuable insight into the four-wave mixing
at the readout. To set the stage we shall first recall the relations
between scattered photons directions. The scattering generated during
the write pulse is emitted around the write beam and falls upon a
circular region of angular diameter $0.5\ \mathrm{mrad}$ in the upper
part of the camera. Similarly, the readout light illuminates the lower
part of the camera of angular diameter $0.5\ \mathrm{mrad}$ placed
in the far field. Both regions are contained in a small central part
of the camera sensor. The angle between the write and read beams equal
to $\theta=2\ \mathrm{mrad}$ inside the $^{87}$Rb cell is precisely
the distance between the centers of both circular regions on the camera.
Due to phase matching illustrated in the Fig. \ref{fig:poziomy-dopasowaniefazowe}(c)
the scattered fields coupled to the same spin-wave with a wave vector
$\mathbf{K}_{b}$ propagate in different directions. They are detected
in the far field on three different pixels of the camera.

For a $\mathbf{K}_{b}$ oriented rightward a pixel located left of
the center in the write region gathers Stokes-coupled light of an
intensity proportional to the number of photons $n_{\mathrm{WS}}$.
Another pair of pixels, right and left of the center in the read region,
gathers anti-Stokes and Stokes coupled light of intensities proportional
to $n_{\mathrm{RA}}$ and $n_{\mathrm{RS}}$ respectively. Note that
the latter pair of pixels also gathers light coupled to the spin-wave
with an opposite wave vector $\mathbf{-K}_{b}$,
and their role is reversed. That is, the left and the right pixels
gather anti-Stokes and Stokes scattering respectively. These contributions
add extra noise when correlated with Stokes scattering $n_{\mathrm{WS}}$
generated in the write-in process.

In order to perform correlation measurements the camera intensifier
was operated in linear regime and long gate pulses were used. We detected
the light scattered during both the write-in and the readout processes.
In each iteration of the experiment we registered entire, randomly
varying camera images. The signal $I_{i}$ at the $i-$th pixel of
the camera was proportional to the intensity of the incoming light.
We calculated the correlation coefficient between intensities of light
registered by any two pixels, $i-$th and $j-$th, of the camera:
\begin{equation}
C_{ij}=\mathrm{corr}(I_{i},I_{j})=\frac{\avg[I_{i}I_{j}]-\avg[I_{i}]\avg[I_{j}]}{\sqrt{\var[I_{i}]\var[I_{j}]}}.\label{eq:corr}
\end{equation}

\begin{figure}[t]
\begin{centering}
\includegraphics[width=0.6\textwidth]{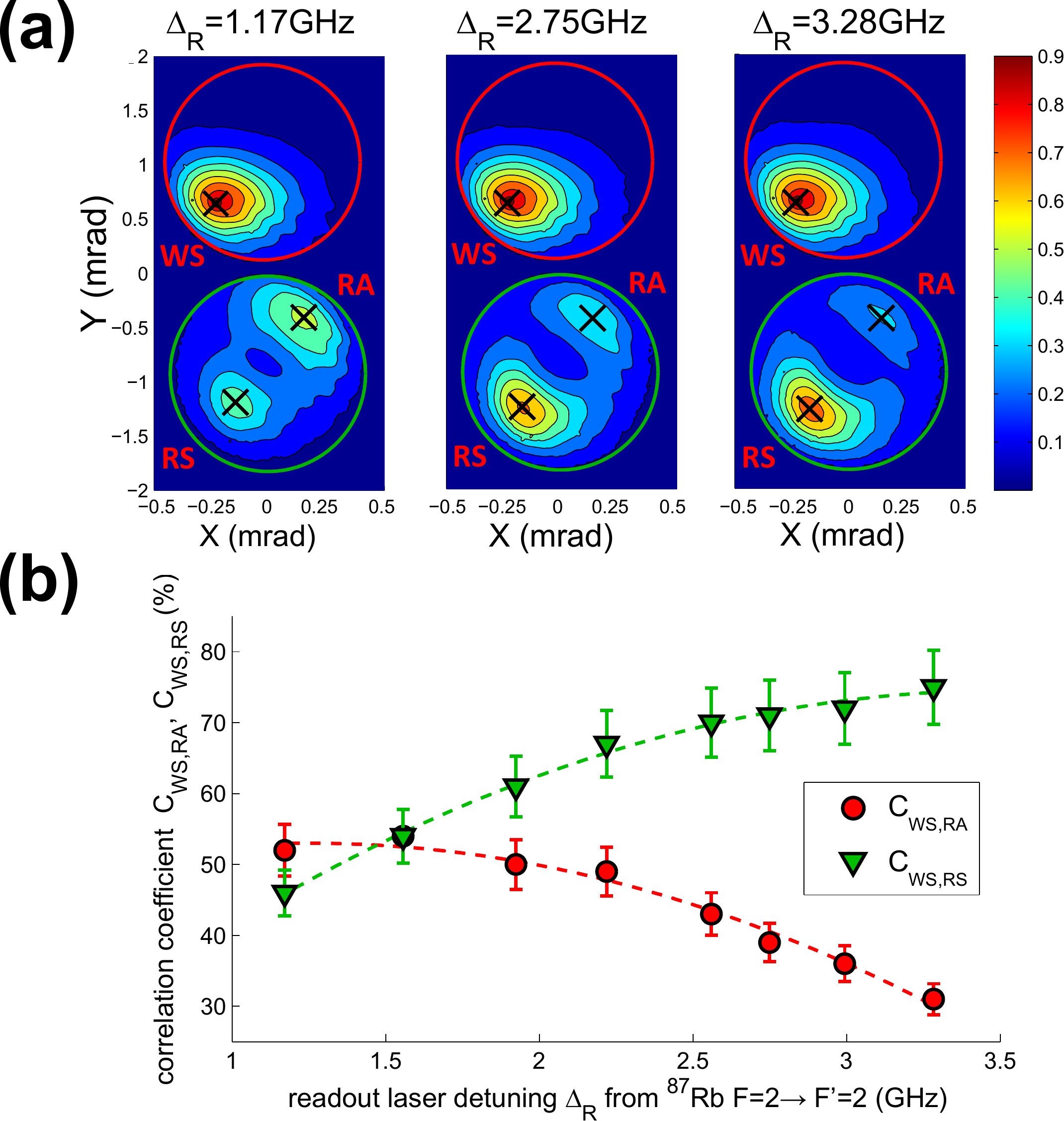} 
\par\end{centering}

\protect\protect\caption{(a) Spatial correlation maps for different read laser detunings from
\protect\protect\textsuperscript{87}Rb F=2 $\rightarrow$ F'=2 resonance.
Red and green circles mark the respective areas where photons scattered
at the write and read stages fell. (b) Correlation coefficients at
the peaks corresponding to anti-Stokes $C_{\mathrm{WS,RA}}$ and Stokes
$C_{\mathrm{WS,RS}}$ versus detuning $\Delta_{R}$ with (quadratic)
trend. \label{fig:mapy_korelacji}}
\end{figure}

In Fig. \ref{fig:mapy_korelacji}(a) we present the maps of correlation
coefficients $C_{ij}$ as in Eq. (\ref{eq:corr}) between one pixel
marked by a cross referred as $i=\mathrm{WS}$ and all other pixels
$j$ in the image. The pixel $i=\mathrm{WS}$ corresponds to the emission
of Stokes field coupled to the spin-wave with the wave vector $|\mathbf{K}_{b}|=45.8\ \mathrm{cm^{-1}}$.
This value was set, so as to spatially resolve anti-Stokes and Stokes
correlation regions. Each correlation map was calculated basing on
$10^{5}$ single-shot images. We limited our calculations to the circles
marked in the maps. They correspond to the regions where the scattering
was registered, centered around the directions of the write and read
beams. In the upper red circles we registered the Stokes scattering
emitted in the write-in process whereas in the bottom green circles
light from readout was observed.

In each map in Fig. \ref{fig:mapy_korelacji}(a) three spots appear.
The maxima of the spots marked by the crosses on pixels $i=\mathrm{WS,\, RA,\, RS}$
correspond to the emission of scattered fields coupled to the same
spin-wave with the wave vector $\mathbf{K}_{b}$ and their positions agree
with the phase matching conditions illustrated in Fig. \ref{fig:poziomy-dopasowaniefazowe}(c).

As we discussed in conclusion of Sec. 2, the correlation between the
write scattered light and two other fields scattered at the readout
stage is mediated by the spin-wave, which caries random number of
excitations $n_{b}$ in each iteration of the experiment. This number
is proportional to the signal on the write pixel $I_{WS}$ and contributes
to the signals $I_{RA}$ and $I_{RS}$. The signals on pixels far
away are correlated to the excitations of other spin-waves, which
vary independently. We verified that correlation maps displayed characteristic
three-spot pattern for arbitrarily chosen reference points except
when two spots in the readout merge. This guarantees that, indeed,
we resolve the Stokes and anti-Stokes scattering in the readout. Correlation
measurements are a practical alternative to spectral filtering in
other experiments with quantum memories \cite{Boyer2008e,Moreau2014}.

Now we can proceed to examine how the anti-Stokes and Stokes scattering
contributions change with the read laser detuning $\Delta_{R}$. In
Fig. \ref{fig:mapy_korelacji}(a) we depicted three correlation maps
for different read laser detunings $\Delta_{R}$. Transition from
the anti-Stokes to Stokes scattering domination can be achieved by
adjusting of the detuning, here from $\Delta_{R}=1.17$ GHz to $\Delta_{R}=3.28$
GHz. To quantify the effect, we calculated the correlation between
the Stokes scattering in write-in and the anti-Stokes scattering ---
$C_{\mathrm{W}\mathrm{S,RA}}$ or the Stokes scattering in readout
--- $C_{\mathrm{W\mathrm{S,RS}}}$. The values of $C_{\mathrm{W\mathrm{S,RA}}}$,
$C_{\mathrm{W\mathrm{S,RS}}}$ were taken as the marked maxima of
the respective anti-Stokes/Stokes regions for various detunings. Fig.
\ref{fig:mapy_korelacji}(b) summarizes the results and illustrates
the transition.

\subsection{Effective gains of anti-Stokes and Stokes readout}

The measured correlation values together with mean intensities and
their variances can be used to extract information about time integrated
readout gains for the anti-Stokes $\bar{G}_{\mathrm{RA}}$ and Stokes
$\bar{G}_{\mathrm{RS}}$ separately. This is accomplished by tracing
the origins of signal observed at respective pixels.

The signal $I_{\mathrm{WS}}$ registered at pixel $i=\mathrm{WS}$
was calibrated to the number of photons. It consists of the part proportional
to the number of photons $n_{\mathrm{WS}}$ generated during the write-in
to the spin-wave of the wave vector $\mathbf{K}_{b}$ and the camera
system read noise $f_{\mathrm{WS}}$: 
\begin{equation}
I_{\mathrm{WS}}=t_{\mathrm{WS}}n_{\mathrm{WS}}+f_{\mathrm{WS}},\label{eq:I_WS}
\end{equation}
where $t_{\mathrm{WS}}$ is the transmission of the write-in Stokes
photons through the filter system. The read noise $f_{\mathrm{WS}}$
is random and its variance depends only on the intensity on the pixel
$\var[f_{i}]\sim I_{i}^{2}.$ We made the calibration of the noise
prior to the measurements, which allowed us to exactly determine the
variance of the read noise at each pixel.

The signal in the readout at pixel $i=\mathrm{RA}$ can be decomposed
into the following contributions: 
\begin{equation}
I_{\mathrm{RA}}=t_{\mathrm{RA}}n_{\mathrm{RA}}+t_{\mathrm{RS}}n'_{\mathrm{RS}}+f_{\mathrm{RA}},\label{eq:I_RA}
\end{equation}
where $n_{\mathrm{RA}}$ is the number of photons generated in the
anti-Stokes readout from the spin wave of the wave vector $\mathbf{K}_{b}$
and $n'_{\mathrm{RS}}$ is the number of photons generated in the
Stokes readout from the other spin wave with the opposite wave vector
$\mathbf{-K}_{b}$, illuminating the same pixel. $t_{\mathrm{RA}}$
and $t_{\mathrm{RS}}$ are respectively the transmissions of the anti-Stokes
and the Stokes photons and $f_{\mathrm{RA}}$ is the camera system
read noise at the pixel $i=\mathrm{RA}$.

We can write an analogous formula for the components of the signal
registered at the pixel $i=\mathrm{RS}$: 
\begin{equation}
I_{\mathrm{RS}}=t_{\mathrm{RS}}n_{\mathrm{RS}}+t_{\mathrm{RA}}n'_{\mathrm{RA}}+f_{\mathrm{RS}}.\label{eq:I_RS}
\end{equation}

In the actual experiment the number of created spin-wave excitations
$n_{b}$ is smaller than the number of scattered photons $n_{\mathrm{WS}}$
due to decoherence at the write-in stage $n_{b}=\eta_{\mathrm{W}}n_{\mathrm{WS}}$.
That factor $\eta_{\mathrm{W}}$ is write-in efficiency. In turn,
the decoherence at the readout stage leads to a limited efficiency
$\eta_{\mathrm{R}}$ during this process. We include those efficiencies
in the formula Eq. (\ref{eq:photon-gain-relation}), which yields
the phenomenological formulas for the number of retrieved photons:
\begin{equation}
n_{\mathrm{RA}}=\eta_{\mathrm{W}}\eta_{\mathrm{R}}\bar{G}_{\mathrm{RA}}n_{\mathrm{WS}}+\check{S}_{\mathrm{RA}},\label{eq:n_RA}
\end{equation}
\begin{equation}
n_{\mathrm{RS}}=\eta_{\mathrm{W}}\eta_{\mathrm{R}}\bar{G}_{\mathrm{RS}}n_{\mathrm{WS}}+\check{S}_{\mathrm{RS}},\label{eq:n_RS}
\end{equation}
where $\bar{G}_{\mathrm{RA}}$and $\bar{G}_{\mathrm{RS}}$ are the
time integrated gains and $\check{S}_{\mathrm{RA}}$and $\check{S}_{\mathrm{\mathrm{RS}}}$
is the spontaneous emission generated at the readout introduced in
the theoretical model. The factors $\eta_{\mathrm{W}}\eta_{\mathrm{R}}\bar{G}_{\mathrm{RA}}$
and $\eta_{\mathrm{W}}\eta_{\mathrm{R}}\bar{G}_{\mathrm{RS}}$ are
the effective gains of the anti-Stokes and the Stokes readouts in
the experiment.

Eqs. (\ref{eq:I_RA})-(\ref{eq:n_RS}) and simple observations that
$\mathrm{corr}(n_{\mathrm{WS}},\eta_{\mathrm{W}}\eta_{\mathrm{R}}\bar{G}_{\mathrm{RA}}n_{\mathrm{\mathrm{WS}}})=1$
and $\mathrm{corr}(n_{\mathrm{WS}},\check{S}_{i})=\mathrm{corr}(n_{\mathrm{WS}},n'_{i})=\mathrm{corr}(n_{\mathrm{WS}},f_{i})=0$
allow us to calculate the effective gains from the measured quantities:
\begin{equation}
\eta_{\mathrm{W}}\eta_{\mathrm{R}}\bar{G}_{\mathrm{RA}}=\frac{t_{\mathrm{WS}}}{t_{\mathrm{RA}}}\frac{C_{\mathrm{WS,RA}}\sqrt{\var[I_{\mathrm{WS}}]\var[I_{\mathrm{RA}}]}}{\var[I_{\mathrm{WS}}]-\var[f_{\mathrm{WS}}]},\label{eq:exp_GRA}
\end{equation}
\begin{equation}
\eta_{\mathrm{W}}\eta_{\mathrm{R}}\bar{G}_{\mathrm{RS}}=\frac{t_{\mathrm{WS}}}{t_{\mathrm{RS}}}\frac{C_{\mathrm{WS,RS}}\sqrt{\var[I_{\mathrm{WS}}]\var[I_{\mathrm{RS}}]}}{\var[I_{\mathrm{WS}}]-\var[f_{\mathrm{WS}}]}.\label{eq:exp_GRS}
\end{equation}
The above formulas can be verified by evaluating the numerators of
the right hand sides.

\begin{figure}[t]
\begin{centering}
\includegraphics[width=0.55\textwidth]{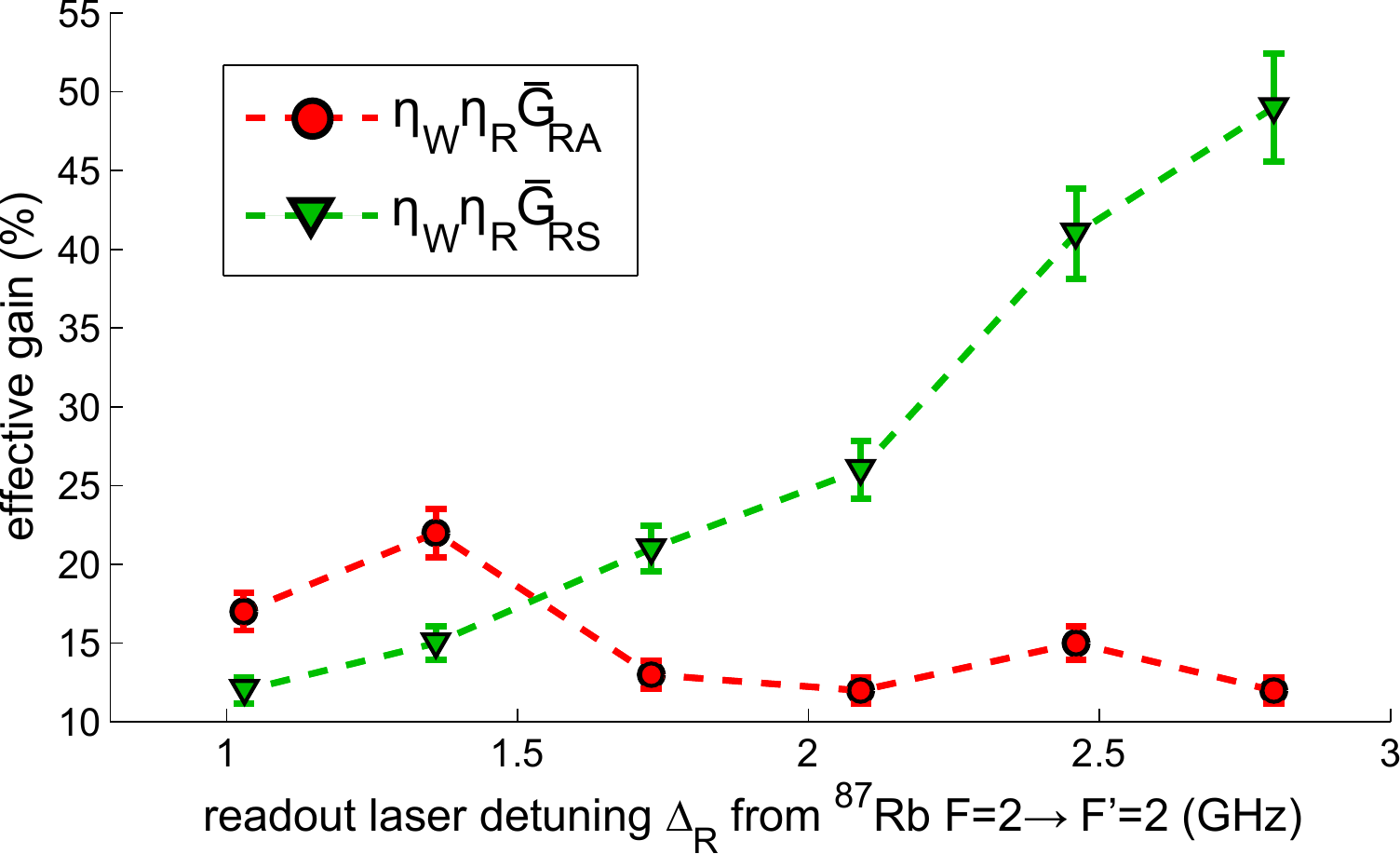} 
\par\end{centering}

\protect\protect\caption{The effective gains of the anti-Stokes scattering $\eta_{\mathrm{W}}\eta_{\mathrm{R}}\bar{G}_{\mathrm{RA}}$
and the Stokes scattering $\eta_{\mathrm{W}}\eta_{\mathrm{R}}\bar{G}_{\mathrm{RS}}$.
For the smallest detuning $\Delta_{R}$ absorption in rubidium cell
starts to contribute.\label{fig:Efficiency}}
\end{figure}

In Fig. \ref{fig:Efficiency} we present the results of the effective
gains of the anti-Stokes scattering $\eta_{\mathrm{W}}\eta_{\mathrm{R}}\bar{G}_{\mathrm{RA}}$
and the Stokes scattering $\eta_{\mathrm{W}}\eta_{\mathrm{R}}\bar{G}_{\mathrm{RS}}$
for different read laser detunings $\Delta_{R}$. The efficiency due
to decoherence in the write-in $\eta_{W}$ is fixed. The efficiency
of the readout $\eta_{R}$ is mostly determined by the decoherece
during this process, thus it also weakly depends on the detuning $\Delta_{R}$.
Thus we expect that the experimentally determined effective gains
vary mostly due to the change of the time integrated gains $\bar{G}_{i}$
and can be compared with the theoretical predictions.

Our results show that in our configuration the Stokes scattering during
readout is an important contribution to the scattered light. The highest
observed efficiency of the anti-Stokes retrieval was $\eta_{\mathrm{W}}\eta_{\mathrm{R}}\bar{G}_{\mathrm{RA}}=$22\%
for $\Delta_{R}=1.37$ GHz whereas the efficiency of the Stokes scattering
$\eta_{\mathrm{W}}\eta_{\mathrm{R}}\bar{G}_{\mathrm{RS}}$ grew with
$\Delta_{R}$. Note that the observed dependence of the anti-Stokes
and Stokes contributions on $\Delta_{R}$ agrees qualitatively with
our simple theoretical model for the integrated gains $\bar{G}_{i}$
presented in Fig. \ref{fig:TeoriaOdOdstrojenia}(d). Note that for
large detunings the anti-Stokes scattering would decay if there was
no Stokes contribution. We suspect that the efficiencies values $\eta_{\mathrm{W}}\eta_{\mathrm{R}}\bar{G}_{\mathrm{RA}}>10\%$
for the anti-Stokes for $\Delta_{R}>2$ GHz originated from the enhancement
due to the full four-wave mixing process. The switch from domination
of anti-Stokes to Stokes scattering occurs at $\Delta_{R}\simeq1.5$~GHz.

\begin{table}[b]
\renewcommand{\arraystretch}{1.0}
\begin{tabular}{|c|c|c|}
\hline 
\multirow{2}{*}{$\Delta_{R}$} & \multicolumn{2}{c|}{$\eta_{\mathrm{W}}\eta_{\mathrm{R}}\bar{G}_{\mathrm{RS}}$}\tabularnewline
\cline{2-3} 
 & $\tau=25\ \mu$s & $\tau=50\ \mu$s\tabularnewline
\hline 
2.0GHz & 18\% & 31\%\tabularnewline
\hline 
3.8GHz & 93\% & 122\%\tabularnewline
\hline 
\end{tabular}\centering

\protect\caption{The effective gains of the Stokes scattering $\eta_{\mathrm{W}}\eta_{\mathrm{R}}\bar{G}_{\mathrm{RS}}$
for different gate durations.  \label{tab:EfficienciesStokes}}
\end{table}

In our setup we were able to manipulate the gate duration and thus
check how the values of the effective gains depend on the time of
interaction. We measured the effective gains for two different gate
durations covering $10\ \mu$s write pulse and a part of $40\ \mu$s
read pulse selected by a gate duration $\tau=25\ \mu$s or $\tau=50\ \mu$s,
as we depicted in Fig. \ref{fig:setup}(b).

We measured the effective gains for two different detunings $\Delta_{R}$
and for the two values of the gate duration $\tau$. We checked that
the anti-Stokes effective gain $\eta_{\mathrm{W}}\eta_{\mathrm{R}}\bar{G}_{RA}=12\%$
remained independent of the interaction time which indicates that
light that has been correlated with the stored spin-waves is always
retrieved in the beginning stage of the readout. On the other hand,
the efficiency of the Stokes scattering varies, as we summarized in
Table \ref{tab:EfficienciesStokes}. As we can see, longer interaction
time increases the effective gain $\eta_{\mathrm{W}}\eta_{\mathrm{R}}\bar{G}_{\mathrm{RS}}$
and in particular we can obtain more correlated light than we did
at the write-in stage $\eta_{\mathrm{W}}\eta_{\mathrm{R}}\bar{G}_{\mathrm{RS}}>100\%$.
This corresponds to the case where the integrated gain $\bar{G}_{RS}$
significantly exceeded unity and overcompensated for the losses,
albeit bringing a large amount of accompanying noise\textbf{ $\bar{S}_{RS}$.}
Such a high gain of the obtained signal can find an application in
various operations with quantum memories. For instance the dominant
Stokes scattering process can be utilized for testing if the memory
is empty\textbf{.} No photons retrieved from the memory at the readout
may suggest that memory was initially empty with a certain probability.
This probability is much higher for amplified readout with Stokes
interaction dominant as opposed to conventional pure anti-Stokes readout
process of limited efficiency.

\section{Conclusions}

In conclusion, we have presented an experimental demonstration of
manipulation of the Hamiltonian in the readout from Raman-type atomic
memory. We measured the temporal evolution of the readout light and
the spatial correlations between the Raman scattering in the write-in
and the readout. Our measurements confirm the adjustability
of coupling parameters corresponding to the anti-Stokes and Stokes
scattering. The results match our simple theoretical model of a full
four-wave mixing process. We resolved the anti-Stokes and the Stokes
scattering contributions to the readout thanks to the phase matching
in the atomic vapor which dictated directional correlations with the
Stokes photons during write-in.

Our results provide a very simple framework for interpretation of
extra noise in experiments on storing light in atomic vapor. When
anti-Stokes scattering is used to map the spin-wave states onto the
states of light, the accompanying Stokes scattering creates unwanted
random photons and atomic excitations. Our results show that, though
inevitable, this contribution can be estimated by our model and perhaps
suppressed by adjusting the coupling light frequency to the other
side of the atomic resonance. There is also an optimal duration for
the anti-Stokes interaction. Beyond the optimum, the spontaneous noise
contribution increases. It may be favorable to switch to noncollinear
configuration where control and quantum fields enter the atomic medium
at a small angle. Then the Stokes scattering photons will become directionally
distinguishable from the anti-Stokes, which may lower the noise in
some experiments.

The design of the Hamiltonian we demonstrated can be implemented in
many types of quantum memories at little or no extra cost. The amplification of readout signal
by Stokes scattering may be very useful in some applications especially
if extra noise is not crucial. This is the case when we use detectors
of small quantum efficiency or we are focused on other properties
of retrieved light e.g. in retrieval of stored images \cite{Firstenberg2013,Shuker2008}.
For instance the amplification in the readout can be utilized as a
robust single-shot projective test to see whether the atomic memory
is in the ground (empty) state. A complete absence of Stokes signal
on an inefficient detector is a relatively rare occurrence if the
extra Stokes gain overcompensates for the losses at the detection
stage. Notably then, complete absence of signal ensures us that the
memory was in the ground state.

Our results can be useful for suppressing unconditional noise floor
in readout of Raman-type quantum memories at the single photon level
\cite{Reim2011,Vurgaftman2013,Michelberger2014}. We provide evidence
for a possibility of engineering a wide range of atom-light interfaces
which can be described theoretically as simultaneous readout and parametric
amplification e.g. quantum non-demolition interaction while coupling
coefficients are equal. The facility to continuously tune the Hamiltonian
coefficients sets the scene for developing new quantum protocols in
room-temperature atomic memories.

\section*{Acknowledgments}

We acknowledge J. Nunn for the insightful discussion and C. Radzewicz
and K. Banaszek for generous support. The project was financed by
the National Science Centre projects no. DEC-2011/03/D/ST2/01941 and
DEC-2013/09/N/ST2/02229.
\end{document}